\documentclass[pra,amssymb,amsfonts,twocolumn]{revtex4}

\def\cA{{\cal A}}

\def\cH{{\cal H}}
\def\cF{{\cal F}}
\def\cL{{\cal L}}
\def\cK{{\cal K}}

\def\ra{\rightarrow}

\def\tr{{\rm tr}}
\def\ftr{field theory}
\def\itr{information theory}
\def\beq{\begin{equation}}
\def\eeq{\end{equation}}
\def\bea{\begin{eqnarray}}
\def\eea{\end{eqnarray}}

\begin{document}

\title{Extension of quantum information theory to
 curved spacetimes}

\author{Daniel R. Terno}
\affiliation{Department of Physics, Technion --- Israel Institute
of Technology, 32000 Haifa, Israel}
\email{terno@physics.technion.ac.il.}

\begin{abstract}

The representation of measurements by positive operator valued
measures and the description of the most general state
transformations by means of completely positive maps are two basic
concepts of quantum information theory. These concepts can be
trivially extended to field theories in curved spacetime if all
the representations of canonical commutation or anticommutation
relations are unitarily equivalent. We show that both concepts can
be applied even when there is no such unitary equivalence.
\end{abstract}

\maketitle

\section{Introduction}

Quantum information theory has many faces and uses variety of
techniques. Among them two concepts are of  outstanding
importance, as can be seen from the standard monographs~[1-6]. The
first concept is a generalization of the elementary representation
of measurements by projection valued measures, associated with
self-adjoint operators (this description is variously labeled as
von Neumann measurement, standard measurement, or PVM). The most
general description of a measurement is given by {\em positive
operator valued measures} (POVM, or generalized measurement).
Next, state transformations are given in terms of completely
positive maps (superoperators), which include the measurement
induced transformation.  An extension of this formalism to  curved
spaces is the subject of this paper.

While many quantum information experiments are performed with
photons, most  discussions are  restricted to non-relativistic
quantum mechanics in general, and finite-dimensional Hilbert
spaces in particular \cite{chu:b}. Undoubtedly this is  sufficient
 for  conventional quantum information
processing. Nevertheless, there is nothing in the mathematical
structure of the theory that prevents its application to objects
that are described by quantum field theories in flat or curved
spaces. Both concepts (generalized measurements and state
transformations) could and should be applied in this context. One
of the initial stimuli to the investigations that led to quantum
\itr~was quantum optics \cite{d:b}. Moreover, C* and W* algebras
are employed both in studies of information dynamics
\cite{d:b,ing} and in field theories \cite{ha:b,ar,wa:q}.

Several aspects of the POVM formalism in relativistic quantum
theory were recently discussed. It was applied to construct
unsharp position observables for field theories in Minkowski space
time \cite{ali,bu}, that were investigated also from the algebraic
point of view \cite{gia}. Causality properties of general state
transformations and measurements were also investigated,
\cite{pe:00,pr:01,pr:02}.

In this paper we show that the POVM formalism and general state
transformations can be successfully applied to quantum field
theories in general curved spacetimes, even in the case when
different Hilbert space representations of canonical commutation
or anticommutation relations are unitarily inequivalent. First we
present some essential facts about generalized measurements and
state transformations. Next, we review how algebraic field theory
deals with the extraction of unambiguous physical predictions from
unitarily inequivalent representations. Finally we use this result
to extend the basic notions of quantum \itr~in general globally
hyperbolic spacetime.

\section{Basics of quantum information}

We begin from the standard mathematical formalism of a measurement
\cite{d:b,ho:b,bu:b}. A PVM description of quantum measurements,
which is only a particular case of generalized measurements in
quantum \itr, is still used in the monographs on quantum \ftr~
\cite{ha:b,ar}. An observable is represented by a self-adjoint
operator $A$ on a Hilbert space $\cH$. Its expectation value on a
state $\rho$ is given by $\tr\rho A$. The spectral theorem
\cite{rs:1} gives a correspondence between self-adjoint operators
and projection valued measures (PVM):
\beq
A=\int_{-\infty}^{\infty} \,\,xP(dx),
\eeq
where $P(\cdot)$ is a PVM. The probability that a measurement of
$A$ on the state $\rho$ will give a result in (the Borel) set $X$
is
\beq
p_{\rho}^A(X)=\tr[\rho P(X)].
\eeq
This description is adequate for the prediction of energy levels,
scattering cross sections and many other physical properties. On
the other hand, joint measurements of conjugate observables,
time-of-arrival measurements, position measurements in
relativistic theory and other detection problems in \itr~require
more general formalism.

The existence of  self-adjoint operators is not an essential part
of the formalism. What is really needed is a method to construct a
probability measure. Thus a generalized description of measurement
by a normalized positive operator valued measure (POVM) is defined
as follows. If $\Omega$ is a (locally compact) set with a
$\sigma$-algebra $\cF$ then a POVM is a map $E:\cF\ra\cL^+(\cH)$
(a map from the set of possible results to the set of positive
bounded operators on the Hilbert space) such that
\beq
E(X)\geq E(\emptyset)=0,
\eeq
for all $X\in\cF$. For a countable collection of disjoint sets in
$\cF$, we have
\beq
E\left(\bigcup_{n=1}^{\infty}X_n\right)=\sum_{n=1}^\infty E(X_n),
\eeq
and
\beq
E(\Omega)=I,
\eeq
while the probability is calculated according to
\beq
p_\rho^E(X)=\tr[\rho E(X)].
\eeq
Therefore POVMs are extensively involved in information problems
in finite-dimensional Hilbert spaces \cite{pe:b,chu:b}. In
relativistic quantum theory they are indispensable for the
construction of reasonable position observables \cite{ali}.
Leaving aside the issue of actual hardware realization of a given
POVM we consider the following important result (Neumark's
theorem).

Let $E$ be a POVM on a Borel $\sigma$-algebra of a compact
metrisable space $\Omega$, $E(X)\in\cL(\cH)$. Then, there is a PVM
$P(\cdot)$ on a Hilbert space $\cK\supset\cH$ such that if $P$ is
the projection of $\cK$ onto $\cH$, then $E(X)$ is the restriction
of $PP(E)P$ to $\cH$ for all Borel sets $X\subseteq\Omega$. For a
separable $\cH$ the space  $\cK$ can be taken separable.

Kraus's `second representation theorem' is based on this result
\cite{krau,bu:b,chu:b}. It guarantees that any POVM can be
realized by adjoining auxiliary system and performing unitary
transformations and standard von Neumann measurements.

The most general state transformation $T(\rho)$ is a normal
completely positive linear map. It can be represented by a
(non-unique) set of bounded operators via the `first
representation theorem' \cite{d:b,krau},
\beq
T[\rho]=\sum_{n\in N}A^\dag_n\rho A_n, \label{trans}
\eeq
where the indexing set $N$ may be taken countable if $\cH$ is
separable. The operators $A_n$ satisfy $\sum A^\dag_n A_n\leq 1$,
with the equality holding when the operation that is represented
by $T$ is non-selective (i.e., accomplished with certainty).

 The only strong mathematical requirement in both representation theorems is a
demand for Hilbert space to be separable. This requirement may
appear problematical, since there is a persistent belief that
quantum \ftr~necessarily implies non-separable Hilbert spaces. It
is not so: this misconception was dismissed on general grounds a
long time ago \cite{pct}. Fock spaces built from one-particle
spaces of free field theories are separable. Actual Hilbert spaces
in various important curved spacetime models, such as Hawking
radiation or Unruh effect, are indeed separable
\cite{wa:q,ha:75,un}.

Transformations that are described by Eq.~(\ref{trans}) form a
semigroup. A Lindblad  equation \cite{l} governs the time
evolution of the states (or in dual picture, of the observables).
Typically in the relativistic theory $\rho$ is the input and
$T[\rho]$--- the output scattering state. Usually $T$  is given by
the $S$-matrix, which is unitary (either rigorously or formally).
Hawking has introduced a non-unitary superscattering operator that
maps density matrices describing the initial situation to density
matrices describing the final situation for the process of black
hole formation and evaporation \cite{ha:s}. Such a step has been
considered controversial in the conventional framework, but it is
quite natural in quantum \itr.

\section{A result in algebraic field theory}

 It is well-known
that different representations of canonical commutation relations
(CCR) or canonical anticommutation relations (CAR) in field
theories lead to  unitarily inequivalent representations
\cite{ha:b,wa:q,fu}. In the case of Minkowski spacetime existence
of preferred vacuum state enables to define a unique Hilbert space
representation. A similar construction is possible in stationary
curved spacetimes \cite{wa:q,fu}. However, in a general globally
hyperbolic spacetime this is impossible and one is faced with
multiple inequivalent representations. The simplest example is the
Unruh effect, where the operators of Bogolybov transformation are
unitary only formally \cite{un,un:w}. The algebraic approach to
the field theory makes it possible to overcome this difficulty.

Algebraic quantum \ftr~is presented in the books of Haag
\cite{ha:b} and Araki\,\cite{ar}, and mathematical exposition of
its results is given in the monograph of Baumg\"{a}rtel and
Wellenberg \cite{ne}. It can be naturally applied to  quantum
\ftr~in curved spacetime \cite{ha:b,wa:q,wa:91}. We mention here
only some simple facts. A basic structure of the theory is an
algebra of local operators (i.e. ones  confined to some open
regions of spacetime), that actually are suitably smeared field
operators. States are normalized positive linear functionals on
the algebra. A representation of an algebra is obtained using
suitably defined inner product via the so-called GNS (Gelfand,
Neumark, Segal) construction. After having an one-particle space
(in the case of free field theories), a Fock space is constructed
in the usual way. Canonical commutation relations lead to
unbounded operators, so to be able to utilize the machinery of
C*-algebras it is customary to consider the unitary Weyl algebra
that is constructed by formal exponentiation of the CCR algebra.

Since not all the observables are included in this algebra, the
set of states  defined by being positive linear functionals of
unit norm is too large. The most obvious (and painful) example is
the expectation value of the quantum energy-momentum tensor: a
renormalization procedures actually lead to a restriction to
`physical' states \cite{wa:q,fu}. This restriction can be given a
rigorous mathematical meaning \cite{wa:q,wa:91}.

The algebraic approach to
 field theory leads to the conclusion that physical predictions
of the theory are independent of the choice of a representation
(in the case of a {\em PVM description} of measurements)
\cite{ha:b,wa:q}. This proof is built in several steps that will
be now outlined. Our claim that both POVM and general state
transformation formalism are independent of representation will be
based on it.

First, it should be noted that any measurement can actually be
performed only with a finite accuracy,  a finite number of
outcomes, and a finite number of times. Suppose that we measure
the value $q$ of the observable $Q$ and among $N$ runs a value
$q_j$ is obtained $n_j$ times. A relative frequency $w_j=n_j/N$ is
used to extract a probability estimate or it is taken at face
value and interpreted as the estimate. Thus the information about
a state $\rho$ can be formulated as \cite{ar,pe:t}
\beq
|p_\rho^Q(q_j)-w_j|<\epsilon_j,
\eeq
for some positive $\epsilon_j$. These inequalities induce a
natural topology on the state space, which is called a `physical
topology' \cite{ar}. Mathematically, they define  a weak-*
topology on the state space and weak topology on the set of
observables \cite{ha:b,ar,rs:1}.

The second important fact is that  CAR C*-algebra and the Weyl
C*-algebra that represents CCR are simple \cite{si}, i.e. do not
contain maximal non-trivial ideals. Therefore,  their
representations are faithful (i.e. have  zero kernel). In this
case a theorem of Fell \cite{fe} says that every positive
functional on the algebra that is associated with one of its
representations is a weak-* limit of finite sums of  positive
functionals associated with any other representation.

Therefore, one arrives to the conclusion that the determination of
a finite number of expectation values of the elements of $\cA$,
each made with finite accuracy, cannot distinguish between
different representations \cite{ha:b,ar,wa:q}. Following Wald
\cite{wa:q} suppose that $(\cH_1,\pi_1)$ and $(\cH_2, \pi_2)$ are
unitarily inequivalent representations of $\cA$. Let
$A_1,\ldots,A_n\in\cA$ and let $\epsilon_1,\ldots \epsilon_n>0$.
Let $\omega_1$ be an algebraic state corresponding to a density
matrix on $\cH_1$. Then there exists a state $\omega_2$
corresponding to a density matrix on $\cH_2$ such that
\beq
|\omega_1(A_i)-\omega_2(A_i)|<\epsilon_i, \forall i. \label{eq}
\eeq
If the algebra under consideration is a von Neumann algebra, it
includes all the projections associated with its self-adjoint
operators \cite{ne} and the desired conclusion easily follows.
  If the algebra is
considered to be only a C*-algebra, the process is more elaborated
but still leads to the same conclusion \cite{wa:q}.

\section{Extension of the formalism}

However, what about using a POVM? The problem of (an approximate)
localization of particles requires it. When there is a preferred
representation of the algebra, or several unitarily equivalent
representations, the generalization is straightforward. What
happens when there is no unitary equivalence? Fortunately, since a
position POVM is constructed from suitably smeared  field
operators and related projections, the  discussion in the previous
section is sufficient. They can be approximated arbitrarily close
by polynomial functions of algebra elements, and thus belong
either to an algebra (von Neumann case) or to its weak closure. In
any case, the choice of the representation remains irrelevant.

In particular, in the case of Unruh effect
 both inertial and accelerated observer would agree on
the localization of the detected particle in the phase space,
i.e., its unsharp position and momenta. A simpler example of such
an agreement is a position detection \cite{un,un:w}. The
corresponding states are thermal ($kT=a/2\pi$) for accelerated
observer and Minkowski vacuum.  The probability calculations (from
first order perturbation theory) can be cast into POVM form. The
POVM consists of two operators,
\beq
E_1=\alpha a^\dag(\chi)a(\chi),~~~~ E_2=1-E_1
\eeq
where $\chi$ is a detector mode function and $\alpha$ is the
probability of detection \cite{un:w}. An application of the formal
unitary operator by ${UE_iU^\dag}$, suitably truncated
\cite{wa:q,un:w}, gives its Minkowskian counterpart.

Considering POVMs in general, a question arises whether we can
give to all the operators on a particular Hilbert space an
invariant meaning. It turns out that this is always possible.

We start from some  representation $(\cH,\pi)$ of the algebra
$\cA$. We adjoin to it the set of all POVM $E(X)$ operators, and
consider the smallest algebra that contains the union of
$\pi(\cA)$ and the set of all POVMs. It is easy to see that this
 algebra is again simple. It is a  representation
of some simple abstract algebra $\tilde{\cA}$, with $\cA$ being a
proper subalgebra.  Then $\tilde{\cA}\setminus\cA$ itself is
described by the action of algebraic states on its elements. We
just invert final steps in the GNS construction of the
representation space \cite{ha:b,ar,wa:91}. For any POVM operator
$E$ on $\cH$ that cannot be expressed as a function of elements of
$\pi(\cA)$  we {\em define} the action of the state
$\omega_\alpha$  by
\beq
\omega(\pi^{-1}(E)):=\tr(\rho_\omega E),
\eeq
where $\rho_\omega$ is a density matrix on $\cH$ that corresponds
to the algebraic state $\omega$ via GNS construction. Thus we
arrive to the minimal algebra that contains the canonical
relations of the theory and all possible observables in it.
 From this point the derivation leading to
Eq.~(\ref{eq}) is straightforward.

Incidentally, we get  an argument in favor of considering an
initial CCR/CAR algebra $\cA$ as a von Neumann algebra. The reason
is that in our POVM analysis we had to expand this initial
algebra. By doing so we have adjoined also all the projections
operators that correspond to self-adjoint operators of $\cA$ (a
PVM is a particular case of POVM).

Now we are able to turn to the question of state transformation.
We show that to any transformation of density matrices on a
Hilbert space we can define an algebraic state transformation.
Since the converse of the statement is obvious, we establish a
correspondence between such transformations on different
representation spaces. We define an algebraic transformation
$\Phi$ via its dual action on the elements of $\cA$, following a
usual practice in Banach spaces \cite{krau,rs:1} :
\beq
 \Phi[\omega](A):=\omega(\Phi^\dag[A]),~\forall A\in\cA.
\eeq
When there is a single Hilbert space a  definition of the adjoint
map $T^\dag$ is given by \cite{krau}
\beq
\tr (\rho T^\dag[B]):=\tr(T[\rho]B),
\eeq
where $B\in\cL(\cH)$. To reconstruct $\Phi$ from its
representation $\pi(\Phi)=T$ on the representation space $\cH$ we
limit ourselves to $B\in\pi(\tilde{A})$. We set
\beq
\omega(\Phi^\dag[A]):=\tr (\rho_\omega T^\dag[\pi(A)]).
\eeq
To ensure that a continuity of $\Phi$ follows from the continuity
of $T$ we have to take $\tilde{\cA}$ to be a von Neumann algebra
\cite{ne}. It is straightforward to check that $\Phi$ has all the
properties required from a valid state transformation (of course,
whether this transformation is compatible with the requirements of
causality is a different question \cite{pe:00,pr:01}).

\section{Summary}
We showed that the basic structures of quantum information theory
can be applied to field theory in curved spacetimes. It gives us a
potentially powerful tool for the analysis of the relations of
information, entropy and black holes. In particular, discussion of
the information loss paradox, superscattering operator, etc., gets
a different perspective.

\section*{ACKNOWLEDGMENTS}

Discussions with Asher Peres and his help in the preparation of
the manuscript are gratefully acknowledged. This work is supported
by a grant from the Technion Graduate School.

\end{document}